\documentclass[preprint,showpacs,preprintnumbers,amsmath,amssymb]{revtex4}
\usepackage{graphicx}

\begin{document}
\title{High-temperature ferromagnetism of $sp$ electrons in narrow
impurity bands: Application to CaB$_6$}
\author{D M Edwards$^{1}$ and M I Katsnelson$^{2}$}
\address{$^{1}$ Department of Mathematics, Imperial College,
180 Queen's Gate, London SW7 2BZ, UK}
\address{$^{2}$Institute for Molecules and Materials, Radboud University of Nijmegen,
Toernooiveld 1, NL-6525 ED Nijmegen, The Netherlands}
\date{\today}

\begin{abstract}
Ferromagnetism with high Curie temperature $T_c$, well above room temperature,
and very small saturation moment has been reported in various carbon and boron
systems. It is argued that the magnetization must be very inhomogeneous
with only a small fraction of the sample ferromagnetically ordered. It is
shown that a possible source of high $T_c$ within the ferromagnetic regions
is itinerant electrons occupying a narrow impurity band. Correlation effects
do not reduce the effective interaction which enters the Stoner criterion
in the same way as in a bulk band. It is also shown how, in the impurity
band case, spin wave excitations may not be effective in lowering $T_c$ below
its value given by Stoner theory. These ideas are applied to CaB$_6$ and
a thorough review of the experimental situation in this material is given.
It is suggested that the intrinsic magnetism of the B$_2$ and O$_2$ dimers
might be exploited in suitable structures containing these elements.

\end{abstract}


\maketitle

\section{Introduction}

There is great interest in the possible existence of
ferromagnetism well above room temperature in materials which have
no transition metal or rare earth metal components. Such high
temperature ferromagnetism has been reported in various carbon
systems such as fullerenes
\cite{fuller1,fuller2,fuller3,JMMMreview,makarovabook} and
graphite \cite{JMMMreview,Esq,graphite} and also in systems such
as CaB$_2$C$_2$ \cite{B2C2_1} and CaB$_6$ \cite{r1,r2} containing
boron. The electrons involved in the ferromagnetism are $s$ and
$p$ electrons, rather than $d$ or $f$. The saturation moment in
these systems is invariably very small and it is generally
believed that the ferromagnetism is associated with defects or
impurities. It is natural to suggest that the ferromagnetism may
merely be due to magnetic impurities such as Fe but this appears
to be ruled out experimentally in some cases \cite{Esq,r3}. It is
also difficult to see how local magnetic moments associated with
dilute Fe impurities could be coupled strongly enough to be
ferromagnetically ordered above room temperature. In fact the same
problem exists for local magnetic moments arising from any other
form of defect. We believe that the most likely source of high
temperature ferromagnetism is itinerant electrons occupying a
narrow impurity band.

In this paper we investigate how itinerant electron ferromagnetism
in a narrow impurity band differs from the usual situation in the
3$d$ band of transition metals. In the next section we discuss the
criterion for ferromagnetism and the magnitude of the Curie
temperature $T_c$ within Stoner theory. It is pointed out that the
size of the effective on-site interaction parameter which appears
in the theory is not limited by the width of the impurity band, as
one might expect naively from Kanamori's \cite{kanamori}
$T$-matrix theory of electron correlation. A formal argument
concerning this point is given in section 3. It is well-known that
in transition metals, and in very weak itinerant ferromagnets such
as ZrZn$_2$, Stoner theory overestimates $T_c$ by a large factor.
Instead $T_c$ is determined by low-lying spin fluctuations which
are not included in Stoner theory \cite{moriya}. In section 4 we
argue that this is not necessarily the case in impurity-band
ferromagnets so that $T_c$ may in fact be close to its Stoner
value.

In all the carbon and boron-based ferromagnets the experimental
situation is quite unclear with many conflicting results. To
illustrate this we take the example of CaB$_6$ and in section 5
many experiments on this system are reviewed. An attempt is made
to build up a picture of the system based on the present theory of
impurity-band ferromagnetism. In section 6 we draw some
conclusions.

\section{Stoner theory of impurity band ferromagnetism}

A condition for ferromagnetism in an itinerant electron system is
the familiar Stoner criterion
\begin{equation}
I_{eff}N\left( E_{F}\right) >1,  \label{i}
\end{equation}
where $N\left( E_{F}\right) $ is the density of one-electron
states per atom per spin at the Fermi level in the paramagnetic
state and $I_{eff}$ is an on-site interaction parameter. This
criterion is not satisfied for most of the 3$d$ metals and for
none of the 4$d$ metals. Even for the ferromagnetic metals Fe, Co
and Ni it is only satisfied by a small margin ($I_{eff}N\left(
E_{F}\right) \geq 1.2$) \cite{herring}. The reason for this is
that the large on-site Coulomb interaction is reduced to an
effective interaction  $I_{eff}\simeq W/5$, where $W$ is the width
of the $d$-band, owing to a correlation effect \cite{kanamori}.
Rather than experience the strong on-site interactions two
electrons  (or holes) avoid coming on the same site as far as
possible and thereby increase their kinetic energy due to
increased spatial confinement. Since $N\left( E_{F}\right) \simeq
5/W$ we have $I_{eff}N\left( E_{F}\right) \simeq 1.$ The situation
is quite different when ferromagnetism occurs in a narrow impurity
band. In this case $N\left( E_{F}\right) \simeq n_{imp}/W_{imp}$ \
where $n_{imp}$ is the fraction of impurity atoms and the width
$W_{imp}$ \ of the impurity band can be very small. Also, even if
the bare interactions for $sp$ electrons are strong enough for the
Kanamori effect to operate, the kinetic energy increase due to
increased spatial confinement of holes is governed by the full
width of the valence band, not by the width of the impurity band.
Thus $I_{eff}\gg W_{imp}$ and this is shown formally in section 3
by considering the $T$-matrix for itinerant electrons in the
impurity band. The Stoner criterion $I_{eff}N\left( E_{F}\right)
>1$ demands that $n_{imp}I_{eff}/W_{imp}>1$ and this can be easily
satisfied if $W_{imp}$ \ is sufficiently small.

In Stoner theory the Curie temperature $T_{c}$ is given by
\begin{equation}
I_{eff}\int dE\left( -\frac{\partial f}{\partial E}\right) N\left(
E\right) =1  \label{ii}
\end{equation}
where $f\left( E\right) =\left\{ \exp \left[ \left( E-\mu \right) /k_{B}T_{c}%
\right] +1\right\} ^{-1},$ $\mu $ is the chemical potential and
$N\left( E\right) $ is the density of states per atom per spin in
the band. For simplicity we consider a half-filled impurity band with
constant density of
states $N\left( E\right) =n_{imp}/W_{imp}$ in the range $%
-W_{imp}/2<E<W_{imp}/2.$ Here it is assumed that each impurity
atom introduces one state of each spin into the band and provides
one electron (or hole). Owing to the assumption of half-filling,
$\mu =0$ at all temperatures so that Eq.(\ref{ii}) gives
\begin{equation}
k_{B}T_{c}=W_{imp}/\left[ 4\tanh ^{-1}\left( W_{imp}/I_{eff}n_{imp}\right) %
\right] .  \label{iii}
\end{equation}
The right-hand side is a monotonic decreasing function of
$W_{imp},$ so letting $W_{imp}\rightarrow 0$ we have
\begin{equation}
k_{B}T_{c}<I_{eff}n_{imp}/4.  \label{iv}
\end{equation}
To obtain a Curie temperature above room temperature, with a
typical value of $I_{eff}=1$ eV, we require  $n_{imp}>0.1.$
Assuming that there is complete spin alignment in the ground
state, which for the rectangular band considered here occurs
whenever the Stoner criterion $n_{imp}I_{eff}>W_{imp}$ is
satisfied, the saturation moment is therefore greater than $0.1\mu
_{B}/$ atom. If there is only partial spin alignment in the ground
state the argument of section 4 supporting applicability of the
Stoner model fails and low energy spin fluctuations are likely to
reduce $T_{c}$ \ far below the Stoner value. We conclude that room
temperature ferromagnetism and a uniformly distributed saturation
moment much less than $0.1\mu_{B}$ per atom are incompatible.

The observed magnetizations in graphite and C$_{60}$ are of the
order $10^{-3}Am^{2}/kg\simeq 2\times 10^{-6}\mu _{B}$ per C atom
but typically an order of magnitude larger in CaB$_{6}$. We must
conclude that the magnetization is very inhomogeneous with perhaps
only a fraction 10$^{-4}$ of the sample ferromagnetically ordered.
Some evidence for this inhomogeneity in CaB$_{6}$ is discussed in
section 5.

\section{The effect of electron correlations}

As we discussed in the previous section, renormalization of the
effective Stoner interaction due to correlation effects is of
crucial importance for understanding  itinerant-electron
ferromagnetism. Formally, this renormalization can be taken into
account via the $T$-matrix approach \cite {tmatr,kanamori} which
is exact for the case of small electron (or hole) concentration
but is qualitatively adequate for arbitrary band filling. We start
with the general many-body Hamiltonian:

\begin{eqnarray}
H &=&H_{t}+H_{U}  \notag \\
H_{t} &=&\sum\limits_{\lambda \lambda ^{\prime }\sigma }t_{\lambda
\lambda
^{\prime }}c_{\lambda \sigma }^{+}c_{\lambda ^{\prime }\sigma }  \notag \\
H_{U} &=&\frac{1}{2}\sum\limits_{\left\{ \lambda _{i}\right\}
\sigma \sigma ^{\prime }}\left\langle \lambda _{1}\lambda
_{2}\left| v\right| \lambda _{1}^{\prime }\lambda _{2}^{\prime
}\right\rangle c_{\lambda _{1}\sigma }^{+}c_{\lambda _{2}\sigma
^{\prime }}^{+}c_{\lambda _{2}^{\prime }\sigma ^{\prime
}}c_{\lambda _{1}^{\prime }\sigma \,,}  \label{hamilU}
\end{eqnarray}
where $\lambda =im$ are the site number $\left( i\right) $ and orbital $%
\left( m\right) $ quantum numbers, $\sigma =\uparrow ,\downarrow $
is the spin projection, $c^{+},c$ are the Fermi creation and
annihilation operators, $H_{t}$ is the hopping Hamiltonian, and
the Coulomb matrix elements are defined in the standard way

\begin{equation}
\left\langle 12\left| v\right| 34\right\rangle =\int d\mathbf{r}d\mathbf{r}%
^{\prime }\psi _{1}^{\ast }(\mathbf{r})\psi _{2}^{\ast
}(\mathbf{r}^{\prime
})v\left( \mathbf{r-r}^{\prime }\right) \psi _{3}(\mathbf{r})\psi _{4}(%
\mathbf{r}^{\prime }),  \label{coulomb}
\end{equation}
where we define for brevity $\lambda _{1}\equiv 1$ etc. Following
Galitskii \cite {tmatr} let us take into account the ladder
($T$-matrix) renormalization of the effective interaction:
\begin{equation}
\left\langle 13\left| T^{\sigma \sigma ^{\prime }}\left( i\Omega
\right)
\right| 24\right\rangle =\left\langle 13\left| v\right| 24\right\rangle -%
\frac{1}{\beta }\sum\limits_{\omega
}\sum\limits_{5678}\left\langle 13\left| v\right| 57\right\rangle
G_{56}^{\sigma }\left( i\omega \right) G_{78}^{\sigma ^{\prime
}}\left( i\Omega -i\omega \right) \left\langle 68\left| T^{\sigma
\sigma ^{\prime }}\left( i\Omega \right) \right| 24\right\rangle ,
\label{tmatrix}
\end{equation}
where $\omega =(2n+1)\pi k_B T$ are the Matsubara frequencies for
temperature $k_B T\equiv \beta ^{-1}$ ($n=0,\pm 1,...$). Using the
spectral representation for the Green's function
\begin{equation}
G_{56}^{\sigma }\left( i\omega \right) =\int\limits_{-\infty }^{\infty }dx%
\frac{\rho _{56}\left( x\right) }{i\omega -x},  \label{spectral}
\end{equation}
substituting Eq.(\ref{spectral}) into Eq.(\ref{tmatrix}) and
calculating the sum over Matsubara frequencies in the usual way
\cite{Mah} one can rewrite
Eq.(\ref{tmatrix}) in terms of the spectral density matrix $\widehat{\rho }%
\left( x\right) :$%
\begin{eqnarray}
\left\langle 13\left| T\left( E\right) \right| 24\right\rangle
&=&\left\langle 13\left| v\right| 24\right\rangle
+\sum\limits_{5678}\left\langle 13\left| v\right| 57\right\rangle
\left\langle 57\left| P\left( E\right) \right| 68\right\rangle
\left\langle
68\left| T\left( E\right) \right| 24\right\rangle ,  \notag \\
\left\langle 57\left| P\left( E\right) \right| 68\right\rangle
&=&\int\limits_{-\infty }^{\infty }dx\int\limits_{-\infty }^{\infty }dy\frac{%
1-f\left( x\right) -f\left( y\right) }{E-x-y}\rho _{56}\left(
x\right) \rho _{78}\left( y\right)   \label{VPT}
\end{eqnarray}
where $E$ is the \textit{real} energy. In this section we neglect
spin polarization since the effective exchange parameter $I_{eff}$
in the Stoner criterion (\ref{i}) should be calculated in the
paramagnetic phase.

If we take into account only on-site Coulomb interaction (the
Hubbard approximation) than the $T$-matrix turns out to be also
diagonal in site indices and the matrix equation (\ref{VPT}) holds
assuming that 1,2,...8 label only orbital indices and
$\widehat{\rho }\left( x\right) $ is local (on-site) spectral
density.

\begin{figure}[tbp]
\includegraphics[width=9cm]{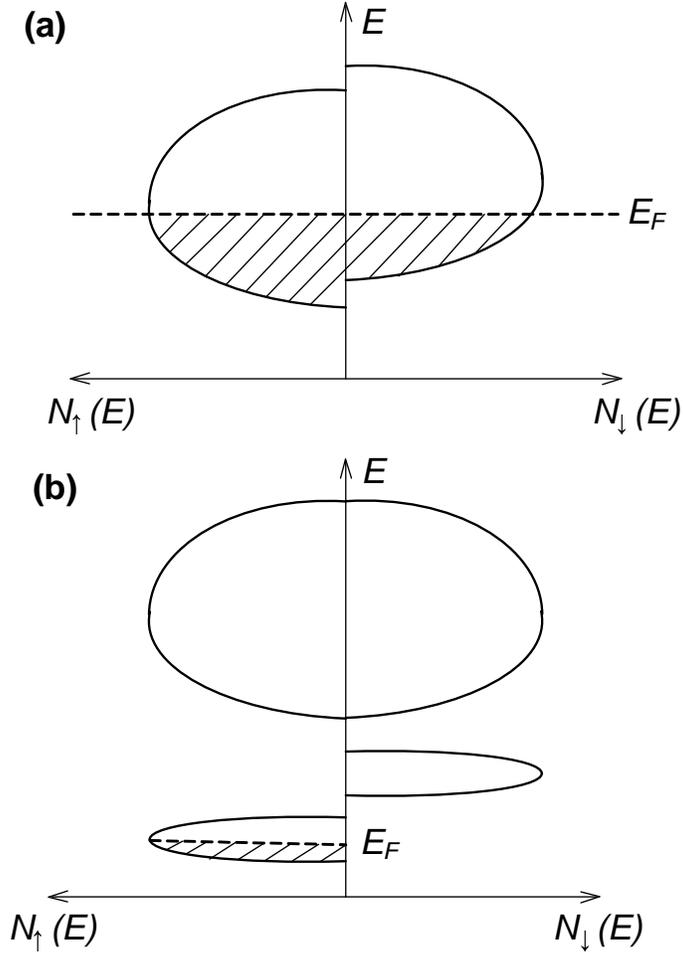}
\caption{Schematic density of states for (a) a weak itinerant
ferromagnet and (b) the ferromagnetic impurity band model}
\label{fig:1}
\end{figure}

For simplicity, we will consider further the case with one orbital
per atom and thus omit orbital indices. The effective interaction
$I_{eff}$ is given by $T(E)$ with $E$ corresponding to the sum of
energies of two occupied electron states \cite{kanamori}. Then
from (\ref{VPT}) we have $I_{eff}=v/(1-Pv)=-1/P$ for large $v$.
Hence in the bulk-band case (Fig. 1(a)) $I_{eff}\simeq W$
\cite{kanamori}. The energy spectrum for the impurity band model
is shown schematically in Fig. 1(b), with a broad main band
(region I) and a narrow impurity band (region II). The figure
shows an impurity band split from the bottom of the main band but
the present considerations are identical for the case of holes in
an impurity band split from the top of a valence band, as in the
model of CaB$_6$ proposed in section 5A. To estimate different
contributions to the function $P(E)$ one should take into account
that
\begin{eqnarray}
\int\limits_{I}\int\limits_{I}\frac{dxdy}{x+y}\rho \left( x\right)
\rho
\left( y\right)  &\sim &\frac{1}{W}Z_{band}^{2},  \notag \\
\int\limits_{II}\int\limits_{II}\frac{dxdy}{x+y}\rho \left(
x\right) \rho
\left( y\right)  &\sim &\frac{1}{W_{imp}}Z_{imp}^{2},  \notag \\
\int\limits_{I}\int\limits_{II}\frac{dxdy}{x+y}\rho \left(
x\right) \rho \left( y\right)  &\sim &\frac{1}{W}\ln \left(
\frac{W}{W_{imp}}\right) Z_{band}Z_{imp},  \label{estim}
\end{eqnarray}
where
\begin{eqnarray}
Z_{imp} &=&\int\limits_{II}dx\rho \left( x\right) ,  \notag \\
Z_{band} &=&\int\limits_{I}dx\rho \left( x\right) =1-Z_{imp}
\label{Z}
\end{eqnarray}
are total spectral weights of the impurity and main bands,
respectively.

To proceed further, one has to specify our model for the impurity
band. Let us consider for simplicity the Slater-Koster model of
single impurity, with non-degenerate band energy $\epsilon \left(
\mathbf{k} \right)$, $\mathbf{k}$ being the quasimomentum, and
on-site only impurity potential $V$ ($V<0$). Then the impurity
site Green's function reads \cite{VK}
\begin{eqnarray}
G_{00}\left( E\right)  &=&\left[ F^{-1}\left( E\right) -V\right]
^{-1},
\notag \\
F\left( E\right)
&=&\frac{1}{N}\sum\limits_{\mathbf{k}}\frac{1}{E-\epsilon \left(
\mathbf{k}\right) }  \label{slater}
\end{eqnarray}
where $N$ is the number of atoms in the crystal. The energy of the
impurity localized state $E_{0}$ and its spectral weight are
determined by the equations
\begin{eqnarray}
VF\left( E_{0}\right)  &=&1,  \notag \\
Z_{imp} &=&\left| \frac{d}{dE}\left[ F^{-1}\left( E\right)
-V\right] \right| ^{-1} _{E=E_{0}}=\frac{1}{V^{2}\left| F^{\prime
}\left( E_{0}\right) \right| }. \label{level}
\end{eqnarray}

For the case of shallow impurity levels
\begin{equation}
\varepsilon \equiv \left| E_{b}-E_{0} \right| \ll W,
\label{shallow}
\end{equation}
where $E_{b}$ is the energy at the bottom of the main band, one
can use the asymptotic forms
\begin{equation}
F\left( E\right) \sim \frac{1}{W} \ln \left( \frac{W}{E_b - E}
\right)  \label{2D}
\end{equation}
and
\begin{equation}
F\left( E\right) \sim  \frac{1}{V_{crit}} + B \frac{\sqrt{E_b -
E}}{W^{3/2}}, \label{3D}
\end{equation}
for two-dimensional (2D) and three-dimensional (3D) cases,
respectively. Here $V_{crit}$ is the critical potential for which
an impurity state splits from the band and $B$ is a dimensionless
constant. We consider the 2D and 3D cases because the top of the
valence band in CaB$_6$ has strong quasi-2D character, as
discussed in section 5A. Substituting Eqs.(\ref{2D}), (\ref{3D})
into Eq.(\ref{level}) one finds
\begin{equation}
Z_{imp}\sim \left\{
\begin{array}{ccc}
W\varepsilon /V^{2}, &  & 2D \\
2W^{3/2}\varepsilon ^{1/2}/V^{2}B, &  & 3D
\end{array}
\right.
\end{equation}
Taking into account Eqs.(\ref{level}),(\ref{2D}) and (\ref{3D})
one can eliminate $V$ and rewrite this estimation in the final
form
\begin{equation}
Z_{imp}\sim \left\{
\begin{array}{ccc}
\frac{\varepsilon \ln ^{2}\left( W/\varepsilon \right) }{W}, &  & 2D \\
\sqrt{\frac{\varepsilon }{W}}, &  & 3D
\end{array}
\right.
\end{equation}
taking $B V_{crit}^2 /2 \sim W^2$. One can see that for the case
of shallow levels the spectral weight of the impurity state is
small. This estimation holds, at least in order of magnitude, also
for many-impurity case when the impurity band is formed. Thus one
can conclude that all contributions (\ref{estim}) to
-$P\left(E\right)$ are much smaller than $1/W_{imp}$ so that
$I_{eff} \gg W_{imp}$. Thus the Kanamori renormalization of the
effective Stoner parameter is not limited by $W_{imp}$ as it is
limited by $W$ in the bulk case. .

\section{Spin waves in the impurity band model}

In Fig. 1 we contrast two types of itinerant electron ferromagnet
with small magnetization in the ground state. Case (a) shows
schematically the situation in a very weak itinerant electron
ferromagnet such as ZrZn$_{2}$. The Stoner criterion is only just
satisfied and a small exchange splitting between the majority
($\uparrow $) and minority ($\downarrow $) spin bands leads to the
Fermi level $E_{F}$ being positioned slightly differently  in the
two bands. Consequently, the majority spin Fermi surface is
slightly larger than the minority spin one and the volume of
$\mathbf{k}$ space between the two Fermi surfaces is proportional
to the ground state magnetization. The Curie temperature is
determined by collective spin fluctuations rather than the single
particle excitations of Stoner theory [13]. Well-defined spin
waves play a negligible role because they are confined to a very
small region of the Brillouin zone. However for sufficiently small
wave-vector $\mathbf{q}$ they exist with energy $Dq^{2}.$ The spin
wave stiffness constant $D$ is small, being proportional to the
magnetization \cite{edw67}.

The situation is different in case (b) where the small
magnetization is associated with complete spin alignment of a low
density of carriers in an impurity band. The dominant low-lying
spin-flip excitations are now long-wavelength spin waves and we
proceed to estimate $D$ in this case. It is shown below that as
long as complete spin alignment is maintained in the ground state
$D$ no longer tends to zero with the magnetization.

For simplicity we consider a one-band model with the disorder
leading to the impurity band treated within the coherent potential
approximation (CPA). We suppose that near the bottom of the band
the Bloch-state energy $\epsilon \left( \mathbf{k}\right) $ in the
absence of disorder and local exchange splitting is given by
\begin{equation}
\epsilon \left( \mathbf{k}\right) =E_{b}+\frac{\hbar
^{2}k^{2}}{2m^{\ast }}. \label{s1}
\end{equation}
Within the CPA the effect of disorder and exchange is contained in
uniform potentials, for an effective medium, which depend on spin
and energy. These
potentials, or self-energies, $\Sigma _{\sigma }\left( E\right) $ for spin $%
\sigma $, would have to be determined self-consistently for a
given detailed model, but here it is sufficient to assume that
they give rise to exchange-split impurity bands as in Fig. 1(b).

The CPA Green's function for spin $\sigma $ is of the form
\begin{equation}
G_{\mathbf{k}}^{\sigma }\left( E\right) =\left[ E-\epsilon \left( \mathbf{k}%
\right) -\Sigma _{\sigma }\left( E\right) \right] ^{-1}
\label{s2}
\end{equation}
with spectral representation (cf Eq.(\ref{spectral}))
\begin{equation}
G_{\mathbf{k}}^{\sigma }\left( E\right) =\int dx\frac{\rho _{\mathbf{k}%
}^{\sigma }\left( x\right) }{E-x+i\eta }=\int dx\frac{\rho _{\mathbf{k}%
,imp}^{\sigma }\left( x\right) +\rho _{\mathbf{k},band}^{\sigma
}\left( x\right) }{E-x+i\eta },  \label{s3}
\end{equation}
$\eta \rightarrow +0.$ Here the spectral functions $\rho _{imp}$
and $\rho _{band}$ refer to the impurity band and main band
respectively.

The spin wave stiffness constant is given by
\begin{equation}
D=\frac{1}{6\pi \left( n_{\uparrow }-n_{\downarrow }\right) }\Im
\int\limits_{-\infty }^{E_{F}}dE\sum\limits_{\mathbf{k}}\left[ G_{\mathbf{k}%
}^{\uparrow }\left( E\right) -G_{\mathbf{k}}^{\downarrow }\left( E\right) %
\right] ^{2}\left| \nabla _{\mathbf{k}}\epsilon \left(
\mathbf{k}\right) \right| ^{2}  \label{s4}
\end{equation}
where $n_{\sigma }$ is the number of carriers of spin $\sigma $
\cite {fukuyama,edwhill}. This result was first derived for
ferromagnetic alloys within the random phase approximation but is
valid within any local approximation, where the self-energy is a
function of energy only \cite {LK,edw2002}.

In the case of complete spin alignment considered here,
$n_{\downarrow }=0$ \ and $G_{\mathbf{k}}^{\downarrow }\left(
E\right) $ is real for $E<E_{F}.$ By using Green's theorem on the
$\mathbf{k}$ sum we find
\begin{eqnarray}
D &=&\frac{1}{6\pi n_{\uparrow }}\Im \int\limits_{-\infty
}^{E_{F}}dE\sum\limits_{\mathbf{k}}\left[
-G_{\mathbf{k}}^{\uparrow }\left(
E\right) \nabla _{\mathbf{k}}^{2}\epsilon \left( \mathbf{k}\right) -2G_{%
\mathbf{k}}^{\uparrow }\left( E\right) G_{\mathbf{k}}^{\downarrow
}\left( E\right) \left| \nabla _{\mathbf{k}}\epsilon \left(
\mathbf{k}\right)
\right| ^{2}\right]   \notag \\
&=&\frac{1}{6n_{\uparrow }}\int\limits_{-\infty }^{E_{F}}dE\sum\limits_{%
\mathbf{k}}\left[ \rho _{\mathbf{k}}^{\uparrow }\left( E\right) \nabla _{%
\mathbf{k}}^{2}\epsilon \left( \mathbf{k}\right) +2\rho _{\mathbf{k}%
}^{\uparrow }\left( E\right) G_{\mathbf{k}}^{\downarrow }\left(
E\right)
\left| \nabla _{\mathbf{k}}\epsilon \left( \mathbf{k}\right) \right| ^{2}%
\right] .  \label{s5}
\end{eqnarray}
Noting that $\nabla _{\mathbf{k}}^{2}\epsilon \left(
\mathbf{k}\right)
/6=\hbar ^{2}/2m^{\ast },$ and using the spectral representation of $G_{%
\mathbf{k}}^{\downarrow },$ we may write
\begin{equation}
D=D_{0}+D_{1}  \label{s6}
\end{equation}
where
\begin{equation}
D_{0}=\hbar ^{2}/2m^{\ast }  \label{s7}
\end{equation}
\begin{equation}
D_{1}=\frac{1}{3n_{\uparrow }}\int\limits_{-\infty }^{E_{F}}dE\sum\limits_{%
\mathbf{k}}\rho _{\mathbf{k}}^{\uparrow }\left( E\right)
\int\limits_{-\infty }^{\infty }dx\frac{\rho
_{\mathbf{k},imp}^{\downarrow
}\left( x\right) +\rho _{\mathbf{k},band}^{\downarrow }\left( x\right) }{E-x}%
\left| \nabla _{\mathbf{k}}\epsilon \left( \mathbf{k}\right)
\right| ^{2}. \label{s8}
\end{equation}
To evaluate the energy integral in $D_{1}$ we approximate the denominator $%
E-x$ in Eq.(\ref{s8}) by $E_{0\uparrow }-x,$ and further approximate it by  $%
-\Delta =E_{0\uparrow }-E_{0\downarrow }$ in the integral involving $\rho _{%
\mathbf{k},imp}^{\downarrow }\left( x\right) .$ Here $E_{0\sigma
}$ is the energy at the centre of the $\sigma $ spin impurity band
and $\Delta =E_{0\downarrow }-E_{0\uparrow }$ is the splitting
between the two impurity bands. Thus $D_{1}$ becomes
\begin{equation}
D_{1}=-\frac{1}{3n_{\uparrow }}\sum\limits_{\mathbf{k}}\left\{ \frac{%
\left\langle n_{\mathbf{k}}^{\uparrow }\right\rangle }{\Delta }\int dx\rho _{%
\mathbf{k},imp}^{\downarrow }\left( x\right) +\left\langle n_{\mathbf{k}%
}^{\uparrow }\right\rangle \int dx\frac{\rho
_{\mathbf{k},band}^{\downarrow
}\left( x\right) }{x-E_{0\uparrow }}\right\} \left( \frac{\hbar ^{2}k}{%
m^{\ast }}\right) ^{2},  \label{s9}
\end{equation}
where $\left\langle n_{\mathbf{k}}^{\sigma }\right\rangle
=\int\limits_{-\infty }^{E_{F}}dE\rho _{\mathbf{k}}^{\sigma
}\left( E\right)
.$ To determine the $\mathbf{k}$ dependence of $\left\langle n_{\mathbf{k}%
}^{\uparrow }\right\rangle $ and $\rho
_{\mathbf{k},imp}^{\downarrow }$ it is sufficient to consider a
single impurity with the on-site potential $V$, as in the previous
section. Then
\begin{equation}
G_{\mathbf{k}}\left( E\right) =\frac{1}{E-\epsilon \left( \mathbf{k}\right) }%
+\frac{t\left( E\right) }{\left[ E-\epsilon \left(
\mathbf{k}\right) \right] ^{2}}  \label{s10}
\end{equation}
with $t\left( E\right) =V/\left[ 1-VF\left( E\right) \right] $
(cf. Eq.(\ref {slater})). For the case of a single impurity the
contribution of the impurity level to the spectral density reads,
similar to Eq.(\ref{level}):
\begin{equation}
-\frac{1}{\pi }\Im G_{\mathbf{k}}\left( E\right) =\frac{1}{\left[
E_{0}-\epsilon \left( \mathbf{k}\right) \right]
^{2}}\frac{-1}{F^{\prime }\left( E_{0}\right) }\delta \left(
E-E_{0}\right) .  \label{s11}
\end{equation}
For a small density of impurities the $\delta $-function broadens
to the shape of the impurity band $\rho _{imp}\left( E\right) $
and we may write
\begin{equation}
\rho _{\mathbf{k},imp}^{\sigma }\left( E\right) =\frac{1}{\left[
E_{0\sigma }-\epsilon \left( \mathbf{k}\right) \right] ^{2}}\rho
_{imp}^{\sigma }\left( E\right) \left\{
\sum\limits_{\mathbf{k}}\frac{1}{\left[ E_{0\sigma }-\epsilon
\left( \mathbf{k}\right) \right] ^{2}}\right\} ^{-1}.  \label{s12}
\end{equation}
Now, $n_{\uparrow }=\int\limits_{-\infty }^{E_{F}}dE\rho
_{imp}^{\uparrow }\left( E\right) $ and
$N_{imp}=\int\limits_{-\infty }^{\infty }dE\rho _{imp}^{\sigma
}\left( E\right) ,$ where $N_{imp}$ is the total number of
impurities and it is assumed that each impurity contributes one
state to the impurity band. Using these equations, together with
Eq.(\ref{s12}), and assuming for simplicity that $E_{0\uparrow
}\simeq E_{0\downarrow }\simeq E_{0}$ (i.e. $\Delta \ll
E_{b}-E_{0}$), we obtain the contribution to $D_{1}$ of the first
term in curly brackets in Eq.(\ref{s9}) in the form
\begin{equation}
-\frac{2\hbar ^{2}}{3m^{\ast }\Delta }N_{imp}\left\{ \sum\limits_{\mathbf{k}}%
\frac{1}{\left[ E_{0}-\epsilon \left( \mathbf{k}\right) \right] ^{2}}%
\right\} ^{-2}\sum\limits_{\mathbf{k}}\frac{\epsilon \left( \mathbf{k}%
\right) -E_{b}}{\left[ E_{0}-\epsilon \left( \mathbf{k}\right)
\right] ^{4}}. \label{s13}
\end{equation}
To estimate the $\mathbf{k}$ sums in this expression we assume a
constant
density of states $C$ per atom in the band $\epsilon \left( \mathbf{k}%
\right) .$ The first term in $D_{1},$ given by (\ref{s13}), the
becomes
\begin{equation}
-\frac{\hbar ^{2}}{2m^{\ast }}\frac{2}{9}\frac{N_{imp}}{I_{eff}n_{\uparrow }C%
}\simeq -\frac{2}{9}\frac{\hbar ^{2}}{2m^{\ast }}  \label{s14}
\end{equation}
if $n_{\uparrow }\simeq N_{imp}$ and $I_{eff}C\simeq 1$.

To estimate the second term in $D_{1}$ we assume that the bulk of
the band
is only weakly perturbed by the disorder so that \thinspace $\rho _{\mathbf{k%
},band}\left( E\right) \simeq \delta \left( E-\epsilon \left( \mathbf{k}%
\right) \right) .$ Combining the result with Eq.(\ref{s14}), and using Eq.(\ref{s7}%
), we obtain
\begin{equation}
D=D_{0}+D_{1}\simeq \frac{1}{9}\frac{\hbar ^{2}}{2m^{\ast }}=\frac{1}{9}%
D_{0}.  \label{s15}
\end{equation}
This is a crude estimate for an oversimplified model. The main
point is that, as long as complete spin alignment is maintained,
$D$ is independent of the ground state magnetization. This
contrasts with the standard case of very weak itinerant electron
magnetism where $D$ is proportional to the magnetization
\cite{edw67}.

A realistic estimate of $D_{0},$ the first term in $D$, may be
obtained within a multi-orbital tight-binding model (e.g.,
Ref.\cite{edw2002}). To make the calculation specific we consider
the case of CaB$_{6}$ where, as discussed in the next section, we
propose that ferromagnetism arises from the complete spin
alignment of holes in an impurity band just above the
valence band. The valence band is assumed to be formed largely from boron 2$%
p$ orbitals and the Hamiltonian is of the form
\begin{eqnarray}
H &=&H_{0}+H_{1},  \notag \\
H_{0} &=&\sum\limits_{\mathbf{k}}\sum\limits_{\mu \mu ^{\prime
}\sigma }V_{\mu \mu ^{\prime }}\left( \mathbf{k}\right) c_{\mu
\mathbf{k\sigma }}^{\dagger }c_{\mu ^{\prime }\mathbf{k\sigma }},
\label{s16}
\end{eqnarray}
where $c_{\mu \mathbf{k\sigma }}^{\dagger }$ creates an electron of spin $%
\sigma $ in a Bloch state of wave-vector $\mathbf{k}$ formed from orbital $%
\mu $. Here $H_{1}$ contains on-site Coulomb and exchange
interactions and an additional one-electron term representing
spin-independent diagonal disorder. For a cubic crystal, such as
CaB$_{6},$
\begin{equation}
D_{0}=\frac{1}{6\left( n_{\uparrow }-n_{\downarrow }\right) }\sum\limits_{%
\mathbf{k}}\sum\limits_{\mu \mu ^{\prime }\sigma }\nabla _{\mathbf{k}%
}^{2}V_{\mu \mu ^{\prime }}\left( \mathbf{k}\right) \left\langle
c_{\mu
\mathbf{k\sigma }}^{\dagger }c_{\mu ^{\prime }\mathbf{k\sigma }%
}\right\rangle   \label{s17}
\end{equation}
where the expectation value is evaluated in the ferromagnetic
ground state. If we assume only nearest-neighbour hopping, and the
on-site orbital energy is taken as zero,  $\nabla
_{\mathbf{k}}^{2}V_{\mu \mu ^{\prime }}\left( \mathbf{k}\right)
=-R^{2}V_{\mu \mu ^{\prime }}\left( \mathbf{k}\right) $ where $R$
is the nearest-neighbour distance. (We neglect the slight
difference in distance between boron atoms within the unit cell of
CaB$_{6}$ and between unit cells). Hence, from Eq.(\ref{s17}),
\begin{equation}
D_{0}=-\frac{R^{2}}{6\left( n_{\uparrow }-n_{\downarrow }\right) }%
\left\langle H_{0}\right\rangle =\frac{1}{12}R^{2}W,  \label{s18}
\end{equation}
where $W$ is the width of the valence band. The last equality follows since $%
\left\langle H_{0}\right\rangle =0$ for a completely full band and $%
n_{\uparrow }-n_{\downarrow }$ holes are removed from an impurity
band formed from states at the top of the band with energy $W/2.$
$D_{0}$ is an upper bound on $D$ since the second term in $D$ is
negative. For CaB$_{6}$ the B-B distance $R=1.7$ \AA  and the
width of the valence band $W=9$ eV \cite{r7}. Hence $D_{0}=2167$
meV$\cdot$ \AA$^{2}$ =3.47$\times 10^{-39}$ Jm$^{2}$. At low
magnon density the magnons may be regarded as non-interacting
bosons with energy $Dq^{2}$. The number of magnons excited per
unit volume at temperature $T$ is then $0.0586\left(
k_{B}T/D\right) ^{3/2}.$ The volume per boron atom in CaB$_{6}$ is
11.88 \AA$^{3}$ so that at $T=600$ K with $D=D_{0}$ the number of
magnons excited per boron atom is $2.57\times 10^{-3}.$ The
corresponding reduction in moment per boron atom is $5.14\times
10^{-3}\mu _{B}$ which is much less than the minimum $T=0$
saturation moment of $0.1\mu _{B}$ discussed in section 2.
Thus if $D=D_{0}$ spin waves are not the dominant excitations controlling $%
T_{c}$ and the Stoner estimate of this quantity should be
realistic. If, however, $D$ is strongly reduced from $D_{0},$ to
$D_{0}/5$ say, the number of magnons excited is an order of
magnitude larger and $T_{c}$ could be reduced considerably from
its Stoner value. Thus to obtain a Curie temperature well above
room temperature we require $D\geq D_{0}/2,$ say. Whether this is
possible can only be decided by realistic calculations of $D$ in a
multiorbital impurity band model. This can certainly be done, but
only if one knows the nature of the impurities or defects
producing the impurity band. No reliance can be placed on the
calculation for a simple one-band model discussed earlier in this
section which gave the estimate $D\simeq D_{0}/9.$

To conclude this section we point out why it should be much easier
to achieve high temperature ferromagnetism in the impurity band
system than in diluted magnetic semiconductors such as (Ga,Mn)As.
In our impurity band model itinerant electron ferromagnetism is
achieved through the high density of states in a narrow impurity
band, which enables the Stoner criterion to be satisfied. In the
diluted semiconductor ferromagnetic order is obtained by an
exchange interaction between the local Mn moments which is mediated
by spin polarized holes in the semiconductor valence band. In this
case Eq.(\ref {s17}) for $D_{0}$ is still valid but the factor
$n_{\uparrow }-n_{\downarrow }$ includes the moment of the Mn
atoms. Thus, for the same number of spin aligned holes in the band
(or impurity band) in both models, and similar bandwidth $W$ and
interatomic distance $R$, the relative value of $D_{0}$ is
controlled by the $n_{\uparrow }-n_{\downarrow }$ factor. Hence in
the diluted ferromagnetic semiconductor $D_{0}$ is reduced by a
factor $M_{band}/\left( M_{band}+M_{loc}\right) $ compared with
the impurity band case. Here $M_{band}$ and $M_{loc}$ are the
contributions to the saturation magnetization of band carriers and
localized moments respectively. With $M_{loc}=4\mu _{B}$ for Mn
atom this factor can be small. The role of spin waves in reducing
the Curie temperature of diluted ferromagnetic semiconductors has
been discussed by Jungwirth et al \cite {Jungw}. The reason why
local moments reduce $D$ so strongly is that they provide no
exchange stiffness, this being entirely due to the band carriers.
It is like hanging weights on a spring without increasing its
stiffness. The oscillation frequency is reduced.

\section{Application to CaB$_6$}

In this section we review most of the experimental data on
CaB$_6$, doped and undoped, and try to build up a theoretical
picture based on the impurity band model. Most other theories are
based on homogeneous models where doping merely changes the number
of electrons in the system. The first such model \cite{r1,r35,r36}
assumes the system behaves as a low-density electron gas which is
predicted to become ferromagnetic for density less than about
$2\times10^{20}cm^{-3}$ ($r_s \geq 20$). Even favourable estimates
of $T_c$ \cite{r35} are far below room temperature. Another early
model \cite{r37,r38,r39,r40} is that of a doped excitonic
insulator which was predicted long ago to be ferromagnetic
\cite{r41}. Problems associated with lack of nesting between
electron and hole Fermi surfaces and with structural instability
have been discussed \cite{r39}. Hotta et al \cite{r42} show that
the excitonic state is unlikely to be stable and also conclude
that ferromagnetism in CaB$_6$ may require the local removal of
cubic symmetry by defects.

Young et al \cite{r1} first observed high-temperature
ferromagnetism with small saturation moment in La-doped CaB$_6$,
SrB$_6$ and BaB$_6$. In Ca$_{1-x}$La$_x$B$_6$ with $x=0.005$ the
saturation moment $M_{s}=3.5 \times 10^{-4} \mu_{B}$/unit cell and
the Curie temperature $T_{c}=600K$. Subsequently Ott et al
\cite{r2} found $M_{s}=2 \times 10^{-4} \mu_{B}$/unit cell and
$T_{c}=900K$ for $x=0.01$. According to the discussion of section
2 the magnetization must be very inhomogeneous with only a
fraction $10^{-4}-10^{-3}$ of the sample ferromagnetically
ordered.

Young et al \cite{r1} found that ferromagnetism does not appear
for $x>0.02$. This has been confirmed recently by Cho et al
\cite{r3} in Ca$_{1-x}$La$_x$B$_6$ crystals with $x=0.03$ and
$0.04$ grown using boron of 99.9\% purity (3N). They did find
ferromagnetism in nominally stoichiometric CaB$_6$, as well as in
Ca$_{1-x}$La$_x$B$_6$ with $x=0.005$, 0.01 and 0.02. All these
ferromagnetic crystals were grown using 3N boron. Cho et al
\cite{r3} found no magnetism in boron-deficient CaB$_6$ (3N) and
in no single crystal of CaB$_6$ (stoichiometric, boron rich or
boron deficient, or La-doped) made using boron of 99.9999\% purity
(6N). Furthermore the temperature-dependent resistivity $\rho(T)$
of CaB$_6$ (6N) exhibits semiconducting behaviour, corresponding
to an energy gap much smaller than the 168 meV deduced by
Vonlanthen et al \cite{r4} from their $\rho(T)$ data, whereas
CaB$_6$ (3N) is metallic with $\rho(0)$ a factor $10^{-4}$ smaller
than in CaB$_6$ (6N). Hall effect measurements on ferromagnetic
undoped CaB$_6$ crystals invariably indicate that electrons are
the dominant carriers \cite{r5}. From the work of Cho et al
\cite{r3} it appears that ferromagnetism is associated with charge
carriers introduced by impurities in 3N boron. The removal of
ferromagnetism by overdoping with La, which adds electrons to the
system, suggests that ferromagnetism is associated with holes in
the valence band, or in an impurity band close to the valence
band. In a ferromagnetic region, with 0.3 aligned hole spins per
unit cell, say, 2\% La doping would remove 0.02 holes per unit
cell which might be enough to move the Fermi level out of a sharp
peak in the density of states so that the Stoner criterion is no
longer satisfied. Clearly if the magnetization were uniform the
critical concentration for uniform La doping in this picture would
be more like 0.002\%. Boron deficiency would also be expected to
remove holes from the valence band. However Hall effect
measurements \cite{r5}, particularly those \cite{r6} on a
ferromagnetic sample with $\rho(T)$ very similar to the CaB$_6$
(3N) crystal, indicate that electrons are present in the
conduction band. This suggests that impurities or defects arising
from 3N boron lead to a small band overlap. The major impurities
in 3N boron were identified as C and Si, with negligible amounts
of magnetic impurities such as Fe \cite{r3}.

\subsection{An impurity band model of ferromagnetism in CaB$_6$}

The first assumption of the model, which seems essential for
reconciling high $T_{c}$ with very small average moment per atom, is the inhomogeneous nature of the
magnetization. Ferromagnetic regions are assumed to have a moment of order 0.1
$\mu_{B}$ per B atom, about 1000 times the average value. As
discussed in the previous section this is consistent with the high
La doping concentration required to suppress ferromagnetism.
Terashima et al \cite{r32} invoke spatially-inhomogeneous
magnetization in Ca$_{0.995}$La$_{0.005}$B$_6$ to try to
understand their susceptibility data. In ferromagnetic resonance
measurements on the same system Kunii \cite{kunii} finds a
magnetization which is three orders of magnitude larger than the
value observed in static measurements. This could be consistent
with our picture.
\begin{figure}[tbp]
\includegraphics[width=9cm]{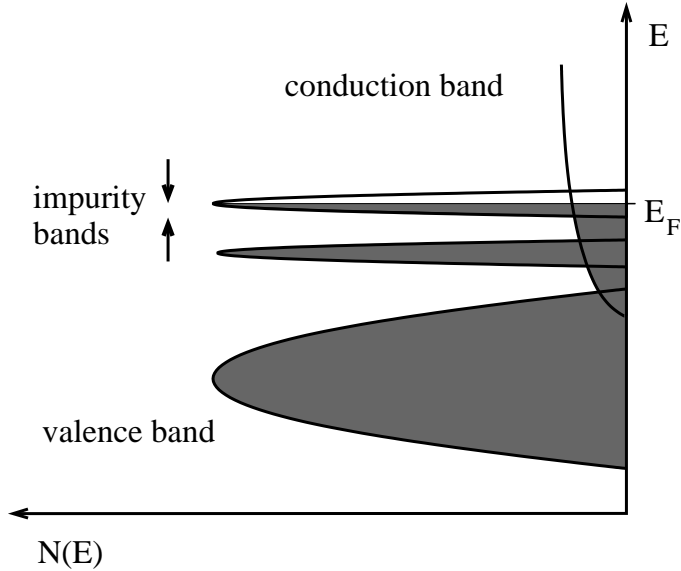}
\caption{Sketch of the proposed electronic structure of
ferromagnetic CaB$_6$.} \label{fig:2}
\end{figure}

The second assumption of the model is that the saturation moment
within a ferromagnetic region arises from complete spin alignment
of holes in a narrow impurity band just above the valence band.
Electrons are also present in an overlapping conduction band whose
states, to a good approximation, do not mix with those of the
valence band. The situation is shown in Fig. 2. Transport is
dominated by the conduction band which may not be much influenced
by the magnetic inhomogeneity of the valence band.

   To proceed further it is necessary to describe
the nature of the CaB$_6$ system. The crystal is cubic, with
Ca$^{2+}$ ions forming a simple cubic lattice. Each cubic unit
cell is occupied by an octahedron of 6 boron atoms, but the
distance between neighbouring B atoms in adjacent cells ($\sim
1.67$ \AA) is considerably less than the distance between B atoms
in one octahedron ($\sim 1.76$ \AA) \cite{r7}. Thus the structure
could be described as a Ca simple cubic lattice with B$_2$ dimers
positioned at the centre of each square face and oriented
perpendicular to that face. Dimerization is a marked effect, the
interatomic spacing in the isolated B$_2$ dimer being $1.59$ \AA,
and it plays a significant role in our theory, as will be seen
later. Clearly the interatomic spacing $2ua$ for dimers in the
lattice can be varied uniformly without changing the cubic
symmetry or the lattice constant $a$. Calculations by Massidda et
al \cite{r7} predict that the band gap in CaB$_6$ reduces by
$0.5eV$ for a 3.5\% decrease in $u$ with constant $a$. The bottom
of the conduction band, at the point $X$ in the zone, falls while
the top of the valence band at $X$ rises. Negative band-gap
corresponds to band overlap. A change from a very small band gap
in the CaB$_6$ (6N) crystals of Cho et al \cite{r3} to band
overlap in CaB$_6$ (3N) might occur by a slight contraction of the
dimers induced by impurities in 3N boron. This impurity effect
would not lead to uniform contraction and an impurity band might
form just above the valence band. This contraction effect might be
produced by many different impurities, so the mechanism leading to
ferromagnetism is not confined to the CaB$_6$ (3N) crystals of Cho
et al \cite{r3}.

The formation of a shallow impurity band is favoured by the
quasi-two-dimensional nature of the band-structure near the top of
the valence band at the $X$ point. All existing band calculations
find a very flat band along $X\Gamma$ which leads to an almost
discontinuous rise in the density of states at the top of the
valence band. However band calculations discussed in section 5B
differ widely in their prediction of the sign and magnitude of the
direct band gap at the $X$ point in CaB$_6$.

The impurity band picture has similarities with the situation in
ferromagnetic transition metals such as Ni and Co. The role of the
transition metal d band is played by a narrow impurity band formed
from boron $p$ orbitals and in both cases there is a broad
overlapping conduction band, this being formed largely from
calcium d orbitals in the CaB$_6$ case. Magnetism in the
transition metal $d$ band is not unexpected since in the isolated
atom the spins of electrons in degenerate d orbitals are aligned
by the Hund's rule mechanism. An analogous situation in the boron
$p$ band becomes clear when one remembers, as discussed above,
that the boron lattice may be regarded as formed from B$_2$
dimers, and that the isolated B$_2$ dimer has a ground state with
total spin $S=1$ \cite{r43, r44}. The two $p$ electrons in B$_2$
occupy degenerate $\pi_x$, $\pi_y$ bonding orbitals, the z axis
being the axis of symmetry of the dimer, and their spins are
aligned according to Hund's rule. The only other magnetic
second-row dimer is O$_2$, the relevant orbitals now being the
corresponding antibonding orbitals.

 \subsection{Further discussion of theory and experiment}

The picture described above is controversial for several reasons.
The first controversial point to consider is whether CaB$_6$ is a
semimetal (semiconductor), with a very small negative (positive)
band gap, or whether it is a semiconductor with a gap of order 1
eV. Following the pioneering band-structure calculation of
Hasegawa and Yanase \cite{r8}, using the local density
approximation (LDA), Massidda et al \cite{r7} found that within
LDA the dimerization effect discussed above led to a small band
overlap. However the difference between small negative and
positive band gap is within calculational error. The top of the
valence band near $X$ is formed from B bonding $p$ orbitals and
the bottom of the conduction band at $X$ is dominated by Ca $d$
orbitals. These states have a different parity on symmetry lines
and the bands therefore cross each other there. Although mixing
occurs at general points in the zone these two bands may therefore
be considered approximately as independent of each other, even
when they overlap. Rodriguez et al \cite{r9} have made similar
calculations for SrB$_6$ and find quite detailed agreement with
optical and transport data \cite{r10} which support the
semimetallic picture. Further strong support for this picture
comes from more recent data \cite{r11} on reflectivity and Hall
resistivity. To obtain a reflectivity which relates to the bulk
band-structure it was necessary to file the surface of the
as-grown crystal to a depth of $10\mu m$ and then polish it.
According to our discussion of Ref.\cite{r3}, CaB$_6$ and SrB$_6$
crystals are expected to be semimetallic if the boron used to make
them is of substantially less than 6N purity. Hall et al
\cite{r12} have made de Haas - van Alphen (dHvA) measurements on
Ca$_{1-x}$La$_x$B$_6$, with $x=0,0.0025,0.005$ and 0.01, and on
SrB$_6$. The La-doped samples were ferromagnetic, except for
$x=0.01$ where in any case no dHvA oscillations were seen owing to
impurity scattering. Orbits attributed to electron and hole
pockets were seen in CaB$_6$ and SrB$_6$ but the hole pockets
dropped out as electrons were added by doping with La. Electron
and hole pockets have also been seen in the related divalent
hexaboride EuB$_6$ by dHvA \cite{r13} and Shubnikov - de Haas
experiments \cite{r14}. The calculations and data so far discussed
in this paragraph are reasonably consistent with the pictured we
proposed in section 5A. However Denlinger et al \cite{r15} report
angle resolved photoemission spectroscopy (ARPES) data, on cleaved
(001) surfaces, showing an $X$-point band gap of about 1 eV for
SrB$_6$ and EuB$_6$. The Fermi level is located just above the
bottom of the conduction band. These authors interpreted the
apparent discrepancy between their ARPES data and the bulk
semimetal picture in terms of an electron-rich surface region,
maybe $60$ \AA  thick, within which the $X$-point band gap varied
rapidly due to variation of the dimerization parameter $u$ between
bulk and surface. They postulated that broken dimer bonds at the
cleaved (001) surface would allow octahedra near the surface to
contract, thus increasing $u$ and consequently the band gap.
Subsequently Denlinger et al \cite{r16,r17} reinterpreted their
large observed gap as characteristic of the bulk. This was largely
in response to a band calculation for CaB$_6$ by Tromp et al
\cite{r18} who used the GW approximation to obtain a bandgap of
$0.8\pm 0.1eV$. The GW approximation generally yields larger, and
more reliable, bandgaps than LDA. Denlinger et al support their
new conclusion with results of X-ray absorption and emission
spectroscopy although the band edges in these data are not
clear-cut. Very similar ARPES data are reported by Souma et al
\cite{r6}. A more recent band calculation for CaB$_6$ using the GW
approximation is that of Kino et al \cite{r19,r20}. They find that
over most of the zone the gap between conduction and valence bands
is enlarged compared with the LDA case, as in Ref.\cite{r18}, but
that a small band overlap nevertheless occurs at the $X$ point.
The subtle reason for this exceptional situation is analysed and
it is suggested that the discrepancy with Ref.\cite{r18} may be
due to an improper treatment of the core contributions in the
pseudopotential plane-wave method by Tromp et al. A similar
pseudopotential method, but with a non-local weighted density
approximation (WDA), was recently used by Wu et al \cite{r21} to
find the same band gap as that of Ref.\cite{r18}. Kino et al
\cite{r19,r20} point out that the difference between a very small
negative and positive band gap is within their calculational
error. The question of semimetal, or small band gap semiconductor,
versus $1eV$ band gap semiconductor in the bulk is clearly not
completely settled. However there is strong evidence for the
former scenario \cite{r11,r12,r19,r20} and we have adopted this
viewpoint here.

The second controversial point we must consider is whether the
small-moment ferromagnetism in these systems arises from spin
alignment of charge carriers within ferromagnetic regions of the
bulk or is associated entirely with magnetic impurities near the
surface. It is also quite possible for these two sources of
magnetism to coexist. The standard method of growing CaB$_6$
crystals uses an Al flux and Otani and Mori \cite{r22,r23}
attribute ferromagnetism to Fe on the surface, the Fe being
introduced as an impurity in the Al flux. They found that
ferromagnetism disappeared completely from crystals of CaB$_6$ and
Ca$_{1-x}$La$_x$B$_6$ after being kept in HCl solution for a few
hours. They conclude that Fe was removed from the surface during
the HCl treatment and that no bulk ferromagnetism exists in these
samples. However many workers have used the same Al flux to grow a
number of crystals, some ferromagnetic and some not. For example
ferromagnetism never exists in overdoped Ca$_{1-x}$La$_x$B$_6$
with $x\geq0.03$ \cite{r1,r3}. Cho et al \cite{r3} found that no
crystal made with 6N boron was ferromagnetic, even though the same
Al flux was used as for ferromagnetic 3N boron samples. However in
some of the latter samples it was found that chemical etching
partially removed the magnetism, although no details are given.
Thus bulk ferromagnetism and surface impurity ferromagnetism may
coexist in these samples. Bennett et al \cite{r5} also find
evidence for such coexistence in a ferromagnetic sample of
CaB$_6$. The saturated moment $M_s$ was reduced by 47\% after a
surface layer of thickness $6000$ \AA was removed by etching, but
no further reduction occurred on removing an additional $6000$
\AA. These workers also found that in 16 different undoped CaB$_6$
crystals there was no correlation between $M_s$ and the electron
density deduced from Hall effect measurements. This is consistent
with our picture that the bulk part of $M_s$ is determined by the
number of holes in the valence band, or an associated impurity
band, whereas the Hall effect is dominated by conduction
electrons. Young et al \cite{r24} report measurements of $M_s$ for
CaB$_6$ grown from Al flux which was deliberately contaminated
with a wide range of concentrations of added Fe. There was no
dependence of $M_s$ on Fe concentration, from which the authors
conclude that alien Fe-B phases are probably not the source of
ferromagnetism in aluminium-flux-grown single crystals. This
contrasts with the behaviour of CaB$_6$ sintered powders for which
Matsubayashi et al \cite{r25} offer convincing evidence that the
observed ferromagnetism is entirely due to FeB and Fe$_2$B phases
on the surface, the Fe originating in the crucible used for
synthesis. Young et al \cite{r24} propose that in
aluminium-flux-grown crystals Fe impurities incorporated in the
bulk are responsible for high-temperature ferromagnetism. It is
assumed that the concentration of Fe is about 0.1 atomic per cent,
being determined by off-stoichiometry of the CaB$_6$. It seems
that such a low concentration of Fe, considered purely as a
magnetic defect, could only lead to ferromagnetism with a very low
$T_c$. In fact Young et al report that inclusion of Co or Ni in
the flux, in addition to Fe, leads to small-moment ferromagnetism
with $T_c < 10K$. They propose that Co and Ni enter CaB$_6$ in
preference to Fe but do not explain why Fe should lead to a $T_c$
100 times larger than is produced by Co or Ni. It is possible that
Fe is one of the defects which lead to a partially occupied
impurity band just above the valence band, whereas Co and Ni do
not. We have argued that such an impurity band can lead to
high-temperature ferromagnetism; this is quite independent of
whether or not the defects responsible for the impurity band are
magnetic impurities. Meegoda et al \cite{r26} argue against any
significant concentrations of Fe in the bulk on the basis of depth
profiling with Auger electron spectroscopy; for a CaB$_6$ crystal
grown in a Al flux they find that Fe impurities are confined to a
surface region with a depth of a few microns.

From the above discussion we conclude that high-temperature
ferromagnetism of CaB$_6$ and Ca$_{1-x}$La$_x$B$_6$ crystals
exists as an inhomogeneous bulk effect. However in as-grown
crystals there is frequently a contribution from Fe impurities in
a surface region which can be eliminated by suitable surface
treatment. The bulk ferromagnetism is clearly associated with
defects or impurities but, as in the case of impurities arising
from use of 3N boron in Ref.\cite{r3}, they need not be
intrinsically magnetic. Jarlborg \cite{r27,r28} looked for
ferromagnetism associated with various defects in SrB$_6$ by means
of LDA band calculations. The point defects were centred in
$2\times2\times2$  or $3\times3\times3$ periodically continued
supercells, corresponding to defect concentrations $x=0.125$ and
0.037 respectively. The point defects considered were La, In and
Al impurities, and a vacancy, replacing one Sr atom. For the La
case a moment of order $0.1\mu_B$ per La impurity was found for
both sizes of supercell. However similar calculations for La in
CaB$_6$ by Monnier and Delley \cite{r29}, using the generalized
gradient approximation (GGA), a variant of LDA, and a
$3\times3\times3$ supercell, resulted in zero moment. Jarlborg
emphasizes that for ferromagnetism it is essential that the Fermi
level falls within an impurity band, which in the case of La is
associated with the conduction band. No magnetism is found in the
In, Al or vacancy case. For In and Al doping with $x=0.125$
Jarlborg finds an almost rigid upward shift of the energy bands
relative to the Fermi level, by about $0.5eV$ in the case of In,
so that holes occupy the valence band. A similar effect also
occurs for the Sr vacancy. For lower doping levels presumably no
rigid shift would occur and the holes might occupy a narrow
impurity band just above the valence band. Jarlborg does not
consider any lattice relaxation around the impurity. Monnier and
Delley \cite{r29} find a large magnetic moment of $2.36\mu_B$
associated with removal of a complete boron octahedron; this
moment is reduced to $1.32\mu_B$ if the B$_6$ octahedron is
replaced by Ca and is lost completely if B$_6$ is replaced by La
or Al. They propose that the large moments associated with B$_6$
vacancies, created during growth due to kinetic effects, are the
source of ferromagnetism. Fisk et al \cite{r30} take up this
suggestion and estimate the density of such vacancies to be
$10^{-4}-10^{-3}$ per unit cell. However no explanation is given
of how such a low concentration of local moments could have a
Curie temperature in the range $600-900K$ (see also
Ref.\cite{r31}).

From the above discussion of Jarlborg's works it appears that Al
substituting for Ca or Sr, or a Ca or Sr vacancy, are possible
sources of the impurity band required for ferromagnetism.
Terashima et al \cite{r32} deduce from a low temperature
resistivity anomaly, also seen by Vonlanthen et al \cite{r33},
that their CaB$_6$ samples were contaminated with Al granules. It
therefore appears that Al does enter the bulk during Al flux
growth and might also substitute for Ca. Hall et al \cite{r12}
find no evidence of Al inclusions in their samples but Al as a
substitutional impurity is presumably not ruled out. However
Monnier and Delley \cite{r29} find that the formation energy of a
Ca vacancy and an Al substitutional impurity are both about $5eV$,
so that the defect concentrations would be negligible if growth
took place under conditions of thermal equilibrium. This is
probably not the case since it is generally believed that
hexaborides, including CaB$_6$, have a tendency to self-doping by
metal vacancies \cite{r2,r4,r14,r32,r33,r34}. It might be that
certain non-magnetic impurities, including those in 3N boron which
are apparently essential for ferromagnetism according to Cho et al
\cite{r3}, facilitate the formation of Ca vacancies or Al
substitutional impurities by kinetic effects during growth, as
suggested by Monnier and Delley \cite{r29} in connection with
B$_6$ vacancies. On the other hand, we have already described how
any impurity leading to more pronounced boron dimerization locally
could lead to the desired impurity band. The effect of Fe
impurities in the bulk on the electronic structure is unknown. The
same can be said of C and Si impurities which are probably
introduced using 3N boron in the work by Cho et al \cite{r3}.

We suggest that there are many ways in which a narrow impurity
band can be formed just above the valence band and we believe that
this is essential for ferromagnetism. The extremely inhomogeneous
nature of the ferromagnetism which we propose might arise in two
ways. One possibility is that the impurities or defects
responsible for the impurity band might be distributed very
inhomogeneously. A second possibility is that the impurity band is
widespread but the Stoner criterion for ferromagnetism is only
satisfied in regions with a particularly favourable impurity
configuration.

\section{Outlook}

A ferromagnetic semiconductor with a Curie temperature $T_c$ well
above room temperature would have great potential for use in
spintronic devices. A well-explored route towards this objective
involves Mn doped III-V compounds but the highest $T_c$ obtained
is well below room temperature. The reason for this failure to
achieve high $T_c$ is discussed at the end of section 4. The
essential point is that although ferromagnetic order is induced by
the large Mn moments they do not contribute to the exchange
stiffness, which is provided by the spin-polarized band carriers.
In fact the Mn moments reduce the spin wave stiffness constant by
a large factor. To achieve high $T_c$ it seems essential that the
band carriers themselves produce ferromagnetic order by their
mutual interaction, without the aid of local magnetic moments.
This requires a high density of states at the Fermi level, to
satisfy the Stoner criterion, and this is most likely to be
achieved in a narrow impurity band. However the estimate of $T_c$
in section 2 indicates that room temperature ferromagnetism is
unlikely unless the number of completely spin-polarized carriers
in the impurity band exceeds 0.1 per bulk atom. It is not clear
whether this can be compatible with a sufficiently narrow impurity
band. Assuming that this scenario is possible, at least in some
regions of the sample, it is argued in section 4 that spin wave
excitation may not significantly reduce $T_c$ from its Stoner
value. If this mechanism is responsible for the ferromagnetism
observed in some carbon systems and in CaB$_6$, the very small
observed magnetization indicates that ferromagnetic order exists
only in a very small fraction of the total volume of the crystal.
Clearly, before any spintronic application can be envisaged, it is
necessary to find a way of producing the conditions for impurity
band ferromagnetism more uniformly throughout the specimen.

In section 5A we pointed out that the CaB$_6$ structure may be
regarded as built from B$_2$ dimers. The isolated dimer is
magnetic, with a moment of 2 $\mu_B$ ; of course this large moment
does not survive in the CaB$_6$ structure where the bonding levels
broaden to form the valence band as electrons hop between bonding
orbitals of adjacent dimers. It would be interesting to find a
boron structure with more pronounced dimerization, and narrower
bands formed from the bonding orbitals, than in CaB$_6$. If such a
structure exists its magnetic properties, doped and undoped, could
be interesting. The same can be said of any structure containing
O$_2$ dimers, linked by hopping between the antibonding orbitals
of the dimer in this case.

{\it Acknowledgement} One of us (DME) wishes to thank J Akimitsu,
M Coey, Z Fisk and H R Ott for useful discussions of
ferromagnetism in CaB$_6$.

\end{document}